\documentclass[aps,prl,twocolumn,showpacs,bibnotes]{revtex4}

\usepackage{graphicx}
\usepackage{amssymb}
\usepackage{bm}
\usepackage{here}

\newcommand{\be}{\begin{equation}}
\newcommand{\ee}{\end{equation}}

\newcommand\pictc[5]{\begin{figure}
                   \centerline{
                   \includegraphics[width=#1\columnwidth]{#3}}
               \protect\caption{\protect\label{fig:#4} #5}
                \end{figure}            }

\newcommand\pict[4][1]{\pictc{#1}{!tb}{#2}{#3}{#4}}
\newcommand\rpict[1]{\ref{fig:#1}}

\newcommand\leqt[1]{\protect\label{eq:#1}}
\newcommand\reqtn[1]{\ref{eq:#1}}
\newcommand\reqt[1]{(\reqtn{#1})}

\newcounter{Fig}

\begin{document}
\begin{sloppy}
\title{Three-dimensional matter-wave vortices in optical lattices}

\author{Tristram J. Alexander}
\author{Elena A. Ostrovskaya}
\author{Andrey A. Sukhorukov}
\author{Yuri S. Kivshar}

\affiliation{Nonlinear Physics Centre and Australian Centre of
Excellence for Quantum-Atom Optics, Research School of Physical
Sciences and Engineering, Australian National University,
Canberra, ACT 0200, Australia}

\begin{abstract}
We predict the existence of spatially localized nontrivial vortex
states of a Bose-Einstein condensate with repulsive atomic
interaction confined by a three-dimensional optical lattice. Such
vortex-like structures include planar vortices, their co- and
counter-rotating bound states, and distinctly three-dimensional
non-planar vortex states. We demonstrate numerically that many of
these vortex structures are remarkably robust, and therefore 
can be generated and observed in experiment.
\end{abstract}


\maketitle

The importance of three-dimensional (3D) vortex configurations
in physics was recognized long ago by Sir W. Thompson when he
suggested that robust vortex rings can be employed to understand
the structure of the atom~\cite{Kelvin:1867:PHM}. Since then
topologically stable vortex lines have been investigated in many
physical contexts, including cosmology, hydrodynamics, optics, and
condensed matter physics. New experimental opportunities for the
study of vortices have appeared in the rapidly developing field
of Bose-Einstein condensates (BECs)~\cite{Anglin:2002-211:Nature}
as condensates provide an excellent test system for the study of
nucleation, dynamics, and the interaction of vortex lines.

One of the outstanding physical problems is the behavior of
vortices in systems with reduced symmetry, such as periodic
structures. BECs loaded into periodic potentials of optical
lattices provide an excellent example of such a system, being
relatively easy to manipulate experimentally.  In a
two-dimensional (2D) geometry, stable vortices localized in an
optical lattice were predicted to exist in the form of {\em
matter-wave gap vortices}~\cite{Ostrovskaya:2004-160405:PRL}.
These vortices are spatially localized states of the condensate
with a  phase circulation that exist due to the interplay of the
repulsive nonlinearity and the anomalous diffraction of the matter
waves due to their Bragg scattering in the lattice. Although 2D
(planar) vortices have also been investigated in 3D discrete
lattice models~\cite{Kevrekidis:2004-80403:PRL+}, nothing is known
about nonlinear dynamics of the vortex states in a fully 3D
continuous lattice geometry.

In this Letter we not only show that planar 2D matter-wave gap vortices persist in the presence of the third lattice dimension, but also predict the existence of {\em completely new vortex states} localized inside the lattice.  In particular we show that the circulation of particles can break the traditional 2D vortex symmetry and form states which circulate simultaneously in {\em multiple symmetry planes}.   Furthermore we show that the 3D lattice provides a unique opportunity to study the {\em coherent interaction of multiple vortices}, and demonstrate that an interacting stack of vortices containing an extended vortex line, which is known to be unstable in homogeneous systems, is {\em stabilized by the lattice}.
As 3D optical lattices provide improved control over condensate excitations~\cite{Greiner:2002-39:NAT}, there is promise for the future experimental study of these highly nontrivial vortex phenomena.

We model the macroscopic dynamics of the Bose-Einstein condensate
in a 3D optical lattice by the rescaled mean-field
Gross-Pitaevskii (GP) equation:
\begin{equation} \leqt{GP}
   i \frac{\partial \Psi}{\partial t}
   + \Delta \Psi
   - V({\bf r}) \Psi
   + \gamma \left|\Psi\right|^2\Psi = 0,
\end{equation}
where $\Psi$ is the complex matter-wave wavefunction, ${\bf r}
\equiv (x, y, z)$, $\Delta \equiv \partial^2/\partial x^2 +
\partial^2/\partial y^2 + \partial^2/\partial z^2$, and
$\gamma=-1$ for repulsive atomic interactions. The characteristic
spatial scale is $a_L = d/\pi$, where $d = \lambda/2$ is the
lattice period, and $\lambda$ is the wavelength of the light
producing the lattice. We consider a cubic lattice potential:
\begin{equation} \leqt{pot}
  V({\bf r}) = V_0 \left( \sin^2 x + \sin^2 y + \sin^2 z \right),
\end{equation}
with the depth $V_0$ scaled by the lattice recoil energy $E_L =
\hbar^2/2ma_L^2$.  The characteristic time scale is given by
$\omega_L = \hbar/E_L$. We have neglected any additional trapping
potential as we are interested in spatial localization due to the
optical lattice and nonlinearity.

We look for the stationary solutions of Eq.~\reqt{GP} using the
ansatz $\Psi({\bf r}, t) = \psi({\bf r}) \exp(-i\mu t)$, where
$\mu$ is the chemical potential. At low atomic densities, the
condensate dynamics  in the lattice is described by the linearized
Eq.~\reqt{GP} where the last term is neglected. Due to the
periodicity of the potential, the corresponding stationary states
can be expressed as a superposition of the Floquet-Bloch modes
satisfying the condition, $\psi_{\bf k}({\bf r}) = B_{\bf
k}({\bf r}){\rm exp}(i{\bf kr})$, where $B_{\bf k}({\bf r}) =
B_{\bf k}({\bf r} + {\bf d})$, and ${\bf k}$ is the
quasi-momentum within the first Brillouin zone of the lattice (see
Fig.~\rpict{bands}). Matter waves can propagate through the
lattice if ${\bf k}$ is real-valued, i.e.~when the chemical
potential of the Floquet-Bloch linear states belongs to one of the
bands (shaded in Fig.~\rpict{bands}). In the 3D
potential~\reqt{pot}, there always exists a semi-infinite spectral
gap extending to $\mu \rightarrow -\infty$, and additional gaps
appear only when the potential depth $V_0$ exceeds the threshold
value $V_{th} \approx 2.24$. In  what follows we choose the value
$V_0 = 6$.
\pict{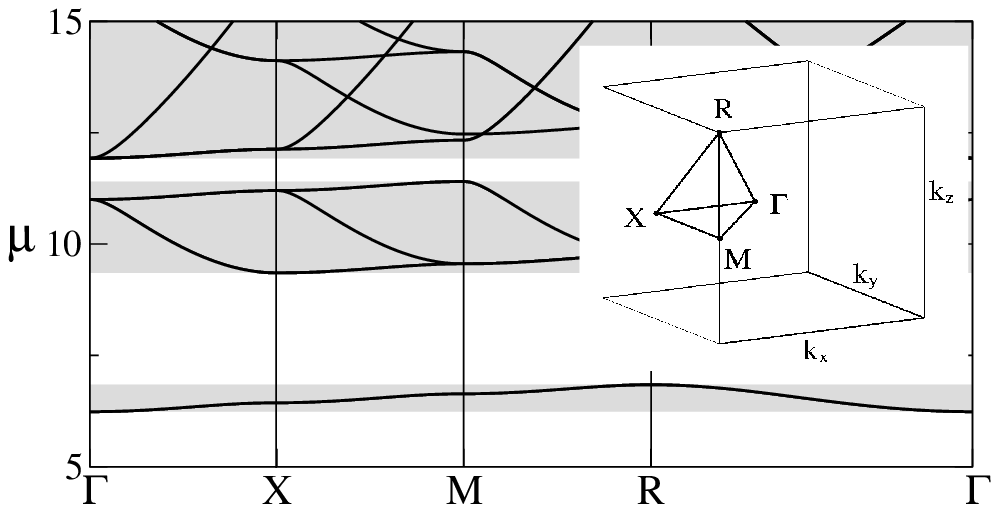}{bands}{Matter-wave spectrum $\mu({\bf k})$
in a three-dimensional optical lattice \reqt{pot} at $V_0 = 6$.
Shaded and open areas show bands and gaps, respectively. Inset
shows the first Brillouin zone and the high-symmetry points in the
reciprocal lattice.}

At higher atomic densities, the effect of repulsive atom-atom
interactions described by the nonlinear term in Eq.~\reqt{GP}
becomes important. In free space, repulsive nonlinearity tends to
accelerate the spreading of the condensate. However, in a lattice
the effective mass of the condensate becomes negative near the
lower gap edges~\cite{Pu:2003-043605:PRA}, allowing for the
formation of localized states inside the spectral gaps~\cite{gap}.
These nonlinear states - {\em gap solitons} - are localized in all
three dimensions. Figure~\rpict{gap} shows an example of a gap
soliton, and the dependence of its norm, $N = \int |\Psi|^2 d{\bf
r}$, on the chemical potential within the first gap calculated
using a numerical functional minimization
technique~\cite{Garcia-Ripoll:2001-1316:SISC}. Despite a
nontrivial phase of the wavefunction [see Fig.~\rpict{gap}(c)],
characteristic of linear Bloch wave tails, there is no particle
flow in the soliton.

Near the lower gap edge the soliton spreads over many lattice
sites and may be described by the free-space GP equation for an
effective envelope of the corresponding Bloch
state~\cite{Pu:2003-043605:PRA}.  In this limit, the 3D gap
soliton is expected to be unstable according to the
Vakhitov-Kolokolov criterion since $\partial N / \partial \mu < 0$
[see Fig.~\rpict{gap}(a)]. Such an instability may result in the
transformation of a weakly localized atomic cloud into a highly
confined state deeper inside the gap [such as shown in
Fig.~\rpict{gap}(b,c)], where $\partial N / \partial \mu > 0$.  We
have investigated the dynamical stability of the strongly
localized gap solitons numerically by introducing $5$\% random
phase and amplitude perturbations.  No evidence of instability can
be seen over the evolution time $t=10^3$, which for ${\rm
^{87}Rb}$ in an optical lattice created with a $0.25 \mu m$ period
is equivalent to ${\rm 17ms}$.

\pict{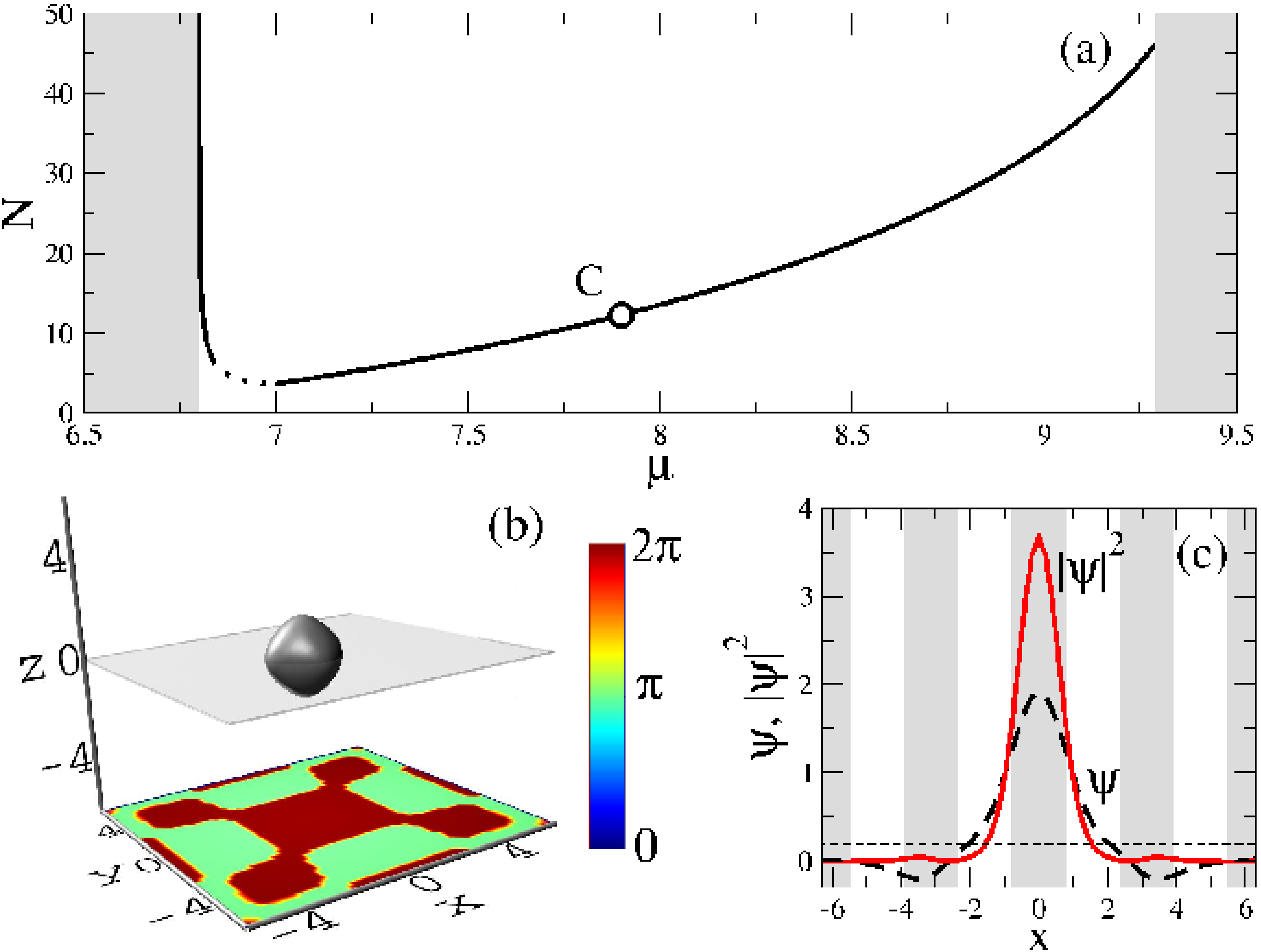}{gap}{(color online) Gap soliton in a 3D optical
lattice with $V_0 = 6$. (a) Norm of the wavefunction $N$ vs. $\mu$ inside the
gap; point C (at $\mu = 7.9$) corresponds to the state shown in
(b) and (c). Line at left band edge calculated in free-space GP limit.  Dotted line an interpolation.  (b) An isosurface of the soliton density, $|\psi|^2$,
at the level $|\psi|^2 = 0.2$, with phase in the $z=0$ plane
(semi-transparent) shown below. (c) The cuts of $\psi$ (dashed)
and $|\psi|^2$ (solid) along the line $y=z=0$. Thin dashed line
corresponds to $|\psi|^2 = 0.2$.  Shaded areas show the lattice minima (parts of the potential less than $V_0/2$).}

Spatially localized stationary states of the repulsive condensate
with a phase circulation can also exist in the gaps of the linear
spectrum. These are localized vortices with stationary density
distribution, but nonvanishing particle flow around a closed path,
which can be characterized by the local current density, ${\bf j}
= {\rm -2Im}( \Psi \nabla_{\bf r} \Psi^\ast)$. The gap vortex states
in a lattice can be thought of as {\em clusters} of gap solitons
$\psi_s$ having different positions on the lattice ${\bf r_m}$ and
phases $\phi_m$. The vortex flow between such solitons occurs due
to a nontrivial phase structure of the full matter-wave function
which is approximately found as $\Psi \simeq \sum_{m=1}^{M}
\psi_s( {\bf r}-{\bf r_m}) \exp(i \phi_m - i\mu t)$. Formation of
a stationary vortex is possible when all the incoming and outgoing
flows are balanced and hence~\cite{Alexander:2004-63901:PRL}:
\begin{equation} \leqt{discrete}
   \sum_{m=1}^{M} c_{nm}{\rm sin}(\phi_m-\phi_n) = 0.
\end{equation}
Each term in this sum defines the flow between gap solitons
numbered $n$ and $m$, which depends on the relative phases and the
nonlinear coupling constants $c_{nm} \equiv c_{mn}
  = \gamma \int \psi_s^3({\bf r}-{\bf r_n})
                \psi_s(  {\bf r}-{\bf r_m}) d {\bf r}
  /        \int \psi_s^2({\bf r}) d {\bf r}$.

The sign of the coupling depends on the sign of the overlap
between the wavefunctions of the gap solitons and the sign of the
nonlinearity. The former may be positive or negative due to the
oscillating tails of the gap solitons [Fig.~\rpict{gap}(c)]. As
the latter is negative ($\gamma = -1$), positive coupling occurs when the
overlapping parts of the wavefunctions are $\pi$ out-of-phase.
Solitons in the cluster can be positioned in the ``nearest
neighbor'' configuration, whereby they are separated by a single
lattice period in the $x$, $y$ or $z$ direction, or in a ``next neighbor''
configuration, whereby the solitons are placed in-plane,
diagonally across a lattice cell (e.g. in the direction $\Gamma
\rightarrow {\rm M}$ in Fig. 1). The ``nearest neighbor'' coupling is
{\em positive}, whereas the ``next neighbor'' coupling is {\em
negative}. Importantly, for a positive coupling, the flow is
directed from a soliton with phase $\phi_m = 0$ to one with $\pi/2$ to $\pi$ etc., while for negative coupling the flow is reversed.  Below we discuss the physical origins of this reversal.

A 3D lattice can support vortex states of different geometries.
First and the most obvious type are the gap vortices localized in
a plane which can be considered as a generalization of 2D gap
vortices~\cite{Ostrovskaya:2004-160405:PRL}. Such 2D states were
shown to have two basic symmetries relative to the lattice sites:
{\em on-site} and {\em off-site}
vortices~\cite{Ostrovskaya:2004-160405:PRL}. The off-site vortex
has a lattice maximum at its center and, therefore, it may consist
of four lobes (atomic density maxima) in a nearest neighbor configuration
[Fig.~\rpict{planar}(a)]. As the nearest neighbor coupling is
positive, the flow circulation is defined in the direction
$0\rightarrow\pi/2\rightarrow\pi\rightarrow 3\pi/2\rightarrow 0$ of the phases of the composite solitons.
The on-site vortices are centered on a lattice site and consist of
four lobes in the next neighbor
configuration [Fig.~\rpict{planar}(b)]. With the nearest neighbor
coupling normalized to $+1$, the on-site vortex has an inter-lobe
next neighbor coupling of $-0.11$, and the particle flow is {\em
opposite} to that expected from the composite soliton phases.  At the physical level, as particle flow is always directed along the phase gradient this seemingly anomalous flow actually arises through the {\em interference of multiple phase singularities} producing a counter-flow opposite to that of the central singularity.

We have examined the dynamical stability of the planar vortices shown in Figs.~\rpict{planar}(a,b) numerically and found that they are stable, at
least up to $t=10^3$. An important consequence of the 3D geometry
is that we may orient these planar vortices in any of the three
symmetry planes of the lattice. Additional planes exist in the
three-dimensional lattice, however they are asymmetric with
respect to the inter-site coupling, and we expect such planes to
host asymmetric planar vortices~\cite{Alexander:2004-63901:PRL}.

\pict{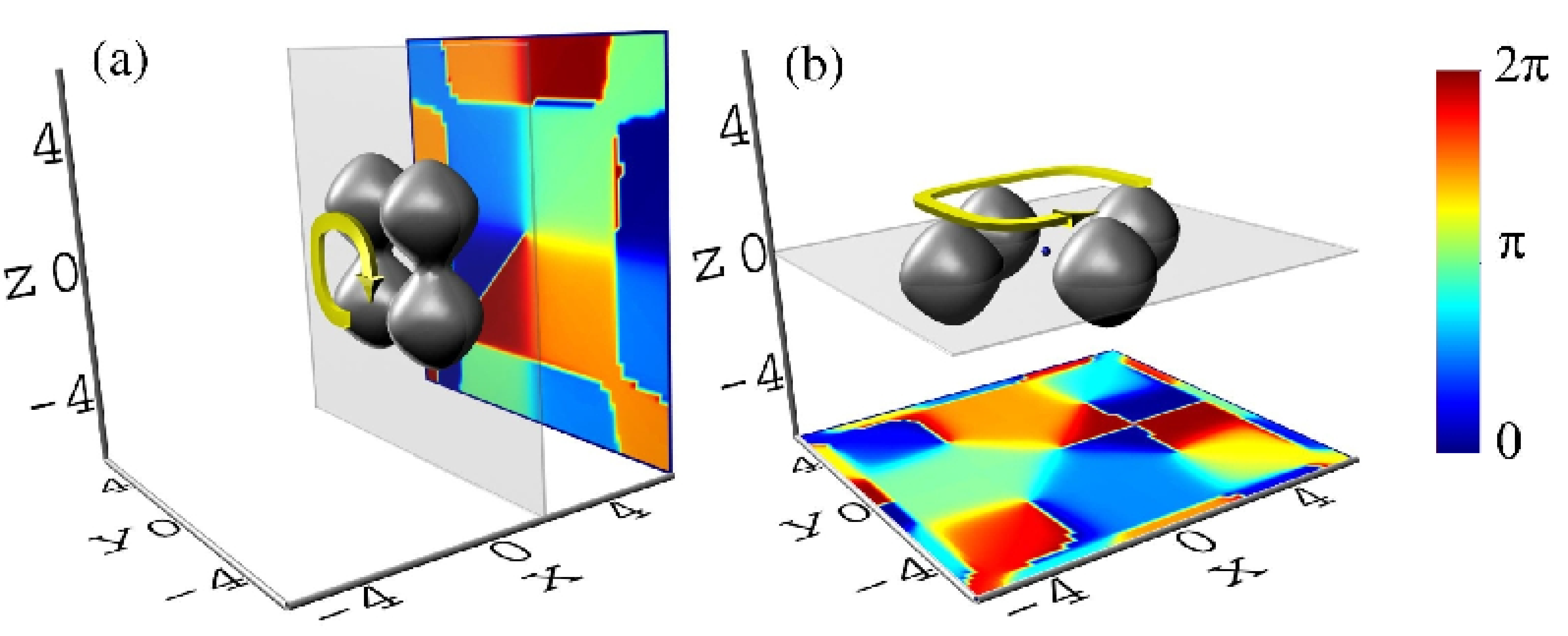}{planar}{(color online) Examples of planar (a) on-site and (b)
off-site vortices ($V_0 = 6$, $\mu = 7.9$).  The orientation in (a) is
the $(x-y)$ plane, while in (b) it is the $(y-z)$ plane.  Arrows
show directions of particle flow.}

A new feature of 3D optical lattices is the possibility of
studying the coherent interaction of {\em multiple planar vortices}.
From the analysis of the discrete equations \reqt{discrete}, we
have identified {\em three basic types} of coupled pairs of planar
vortices: in-phase, out-of-phase, and hybrid-phase.
These may be described by the phases of the
vortex lobes. The in-phase pair has a phase structure, in multiples of
$\pi$, corresponding to
$([0,0],[\frac{1}{2},\frac{1}{2}],[1,1],[\frac{3}{2},\frac{3}{2}])$
where we have used the notation of $[{\rm vortex_1,vortex_2}]$ for
each of the four lobes.  The out-of-phase pair has a phase of
$([0,1],[\frac{1}{2},\frac{3}{2}],[1,0],[\frac{3}{2},\frac{1}{2}])$,
while the hybrid vortex pair has a phase structure
$([0,0],[\frac{1}{2},\frac{3}{2}],[1,1],[\frac{3}{2},\frac{1}{2}])$.
In an equivalent hybrid state the second vortex is shifted in
phase by $\pi$.  The particle flows of the in-phase and
out-of-phase vortices are co-rotating, and hence the pair shares a
singular vortex line at the center [Fig.~\rpict{interacting}(a)].
In contrast, the hybrid vortex states are equivalent to
interacting {\em counter-rotating} vortices. Their vortex lines
have opposite charge and {\em collide}, producing two additional
vortex lines splintering away from the interaction area [see
Fig.~\rpict{interacting}(b)], but conserving the total vortex
charge.  The algebra of these vortex lines and the exciting
possibility of controlling their direction 
will be discussed elsewhere.

Stability of the planar vortex pairs depends critically on the
coupling between vortices. We find no examples of stable
out-of-phase co-rotating and the counter-rotating pairs, whereas
the in-phase co-rotating vortices can be very robust. The in-phase
configuration, due to the positive coupling between
nearest-neighbor lobes, is equivalent to a $\pi$-phase jump
between the wavefunctions, due to the nature of the linear Bloch
waves.  Such phase ``twisting'' has a stabilizing effect as
previously discovered in the framework of discrete nonlinear
systems with attractive
nonlinearities~\cite{Kivshar:1992-3198:PRA}. The actual region of
stability for the co-rotating vortex states depends on the
strength of the coupling (which depends on the values of $\mu$
and $V_0$), in agreement with the results for 2D vortices with an
attractive nonlinearity~\cite{Yang:2004-47:NJP}.  The state shown
in Fig.~\rpict{interacting}(a) is dynamically stable to $t=10^3$.

\pict{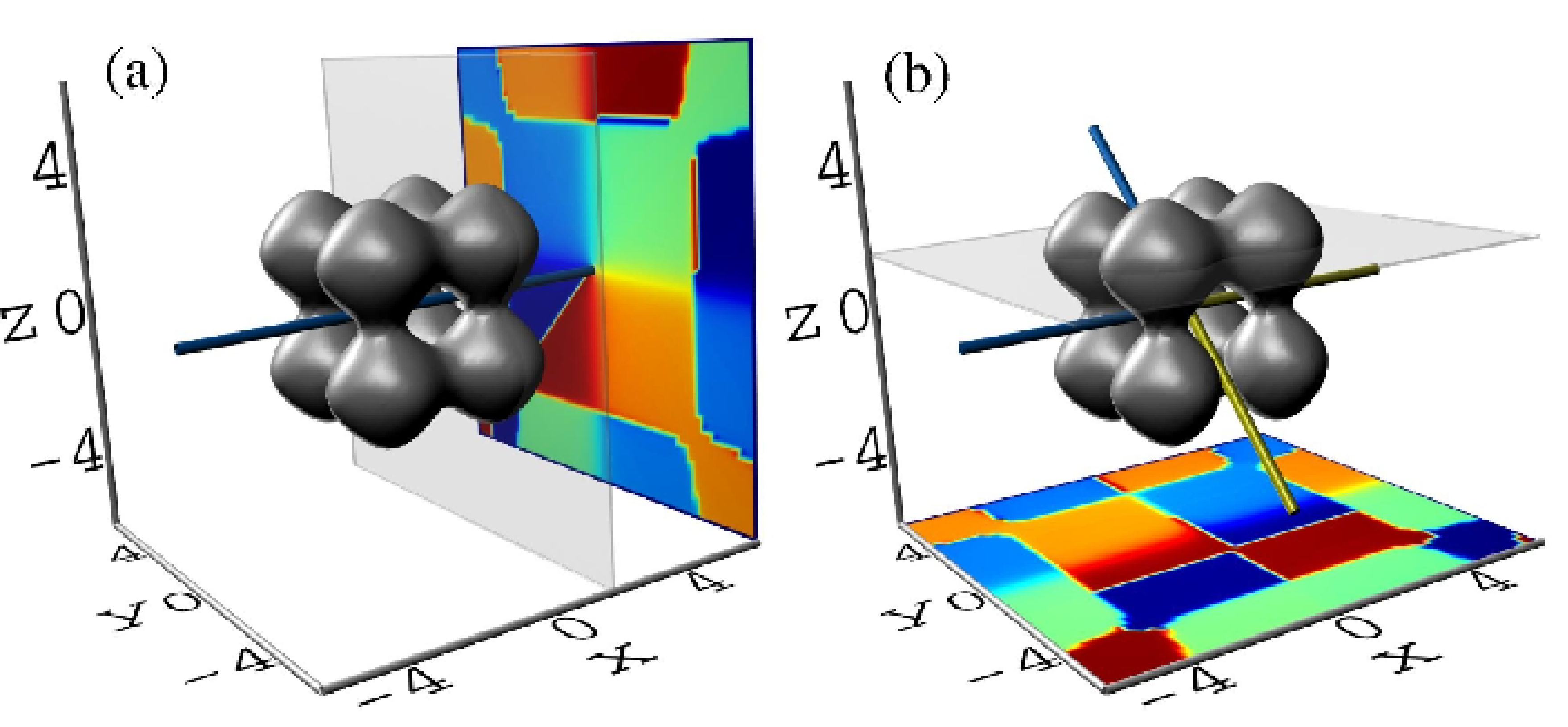}{interacting}{(color online) Examples of
interacting vortex solitons at $V_0 = 6$, $\mu = 7.9$.(a) In-phase
co-rotating vortex pair. (b) Counter-rotating vortex pair.  Lines correspond to primary vortex lines, with different shadings showing opposite topological charges.}

Most remarkably, we find that vortex ``stacks'' of multiple
co-rotating vortices (somewhat analogous to the stacks of
``pancakes''~\cite{Martikainen:2004-70402:PRL}), forming an
extended discrete vortex line, can also be highly robust, in both
the off-site [Fig.~\rpict{stacks}(a)] and on-site configurations
[Fig.~\rpict{stacks}(b)].  These results demonstrate that a 3D
optical lattice can provide transverse stabilization of vortex
lines. This effect can be explored experimentally, since without
the transverse periodicity, vortex lines exhibit strong
`snake-type' instabilities \cite{Martikainen:2004-70402:PRL}. The
possibility of long wavelength modulational instability can not be
ruled out however, especially at long time
scales~\cite{Darmanyan:1999-5994:PRB}. Our numerical simulations
performed with stacks up to 16 lattice sites long have found no
trace of such an instability at $t=10^3$ [Fig. \rpict{stacks}(a)].
When the instability is present, e.g. due to stronger
inter-lobe coupling, it develops rapidly, leading to the
spectacular ``explosion'' of the vortex line [see
Fig.~\rpict{stacks}(b)].

\pict{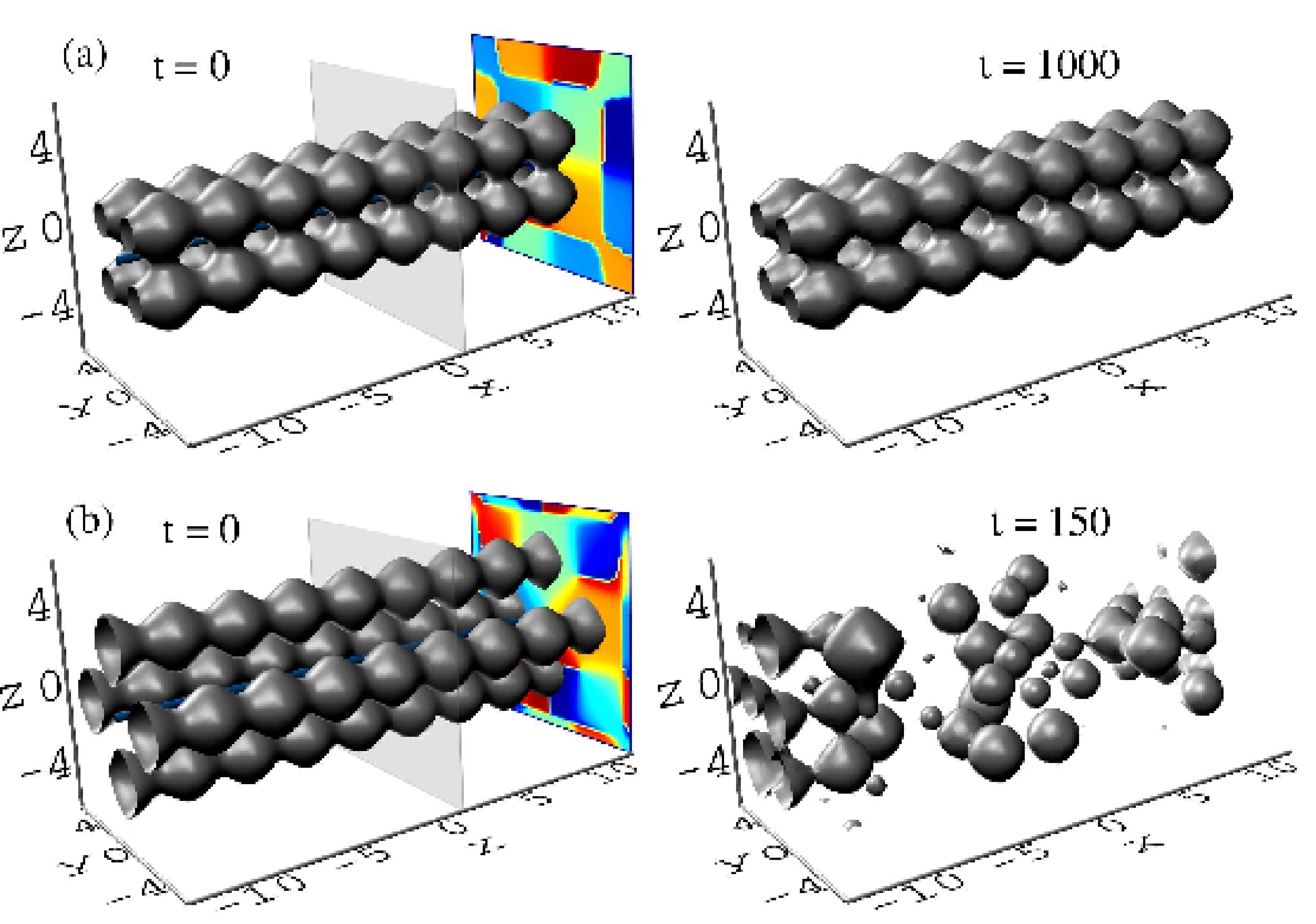}{stacks}{(color online) Example of a stable off-site vortex stack
at $t=0$ (a) and at $t=10^3$ (b) ($\mu = 7.9$).
(c) Unstable on-site vortex stack at $t=0$ and $t=150$ ($\mu=7.5$)(d).}

Apart from the generalizations of planar 2D gap vortices, 3D
lattices can support vortices which have {\em no 2D analogs}.  The
``truly'' 3D vortex with the strongest coupling between lobes is
the four-site state with simple phases
$(0,\frac{1}{2},1,\frac{3}{2})$ (in multiples of $\pi$) shown in
Fig.~\rpict{novel}(b). We term this configuration a ``folded
vortex'', which can be viewed as a combination of two dipoles
$(0,1)$ and $(\frac{1}{2},\frac{3}{2})$ that are not equivalent in
inter-site coupling strengths. Similar coupling structure also
occurs for a planar rhomboid
vortex~\cite{Alexander:2004-63901:PRL}. The simplest fully
symmetric 3D vortex is that of the out-of-plane coupled dipole
state of Fig.~\rpict{novel}(c).  As can be seen by the particle
flow arrows, the vortex circulation is highly nontrivial.

Fully 3D lattices also offer us the possibility of combining
different symmetry elements of 2D vortex states.  The simplest
example is the ``vortex diamond'' shown in Fig.~\rpict{novel}(d),
which consists of two coupled planar vortices lying in the {\em
orthogonal symmetry planes}.  Unlike the incoherently coupled
out-of-plane vortices~\cite{Kevrekidis:2004-80403:PRL+}, this is a
{\em coherent fully 3D vortex} characterized by a single chemical
potential, and it can not be reduced to 2D equivalents.

A remarkable feature of the uniquely 3D gap vortices is their
dynamical stability up to at least $t=10^3$ for a range
of chemical potentials and lattice depths which is much broader than
that of the bound states of two or more planar vortices.  The
experimental generation of these states is an open problem.
However they appear to be an extremely robust means of introducing
localized vorticity into the 3D lattice.

\pict{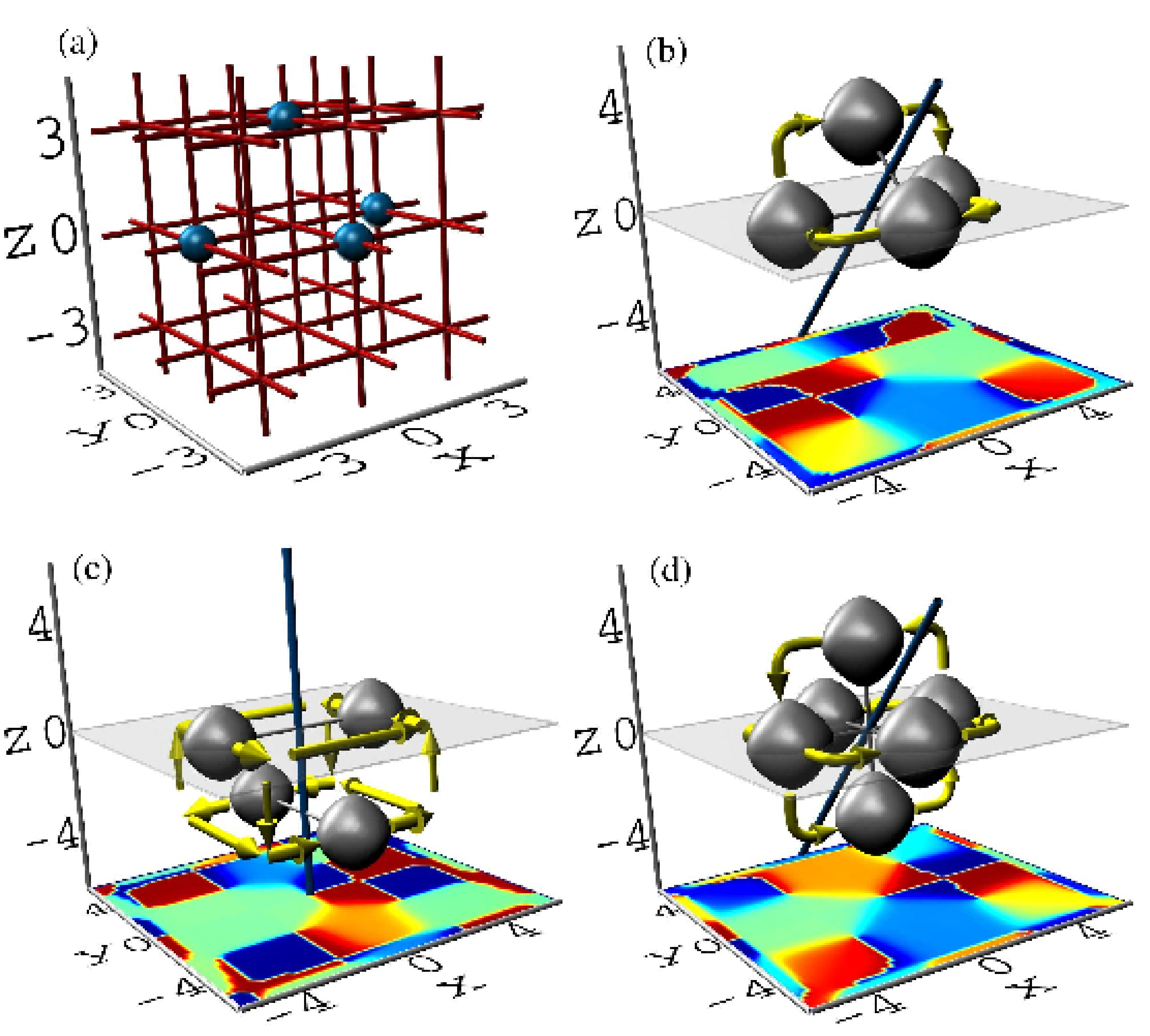}{novel}{(color online) Novel 3D vortices at $V_0 =
6$, $\mu = 7.9$. (a) Folded vortex shown in (b) relative to the
positions of the lattice minima. (c) Crossed-dipole vortex and (d)
vortex diamond.  Thick lines correspond to the primary vortex
lines.}

In conclusion, we have predicted novel classes of spatially
localized vortex structures in BECs loaded into 3D optical
lattices. Such structures include planar vortices, vortex stacks,
folded vortices and vortex diamonds. We have demonstrated
numerically that many of these vortex structures are remarkably
robust, and therefore they may be generated and observed in
experiment. Our results open up the possibility of studying more
complex condensate structures such as localized vortex rings
and knots in periodic potentials.

\end{sloppy}
\end{document}